# The Electronic Properties of Phosphorene/Graphene and Phosphorene/Hexagonal Boron Nitride Heterostructures

Yongqing Cai[†], Gang Zhang[*,†], and Yong-Wei Zhang[*,†]

[†]Institute of High Performance Computing, A*STAR, Singapore 138632

**ABSTRACT:** Vertical integration of two-dimensional materials has recently emerged as an exciting method for the design of novel electronic and optoelectronic devices. Using density functional theory, we investigate the structural and electronic properties of two heterostructures, graphene/phosphorene (G/BP) and hexagonal boron nitride/phosphorene (BN/BP). We found that the interlayer distance, binding energy, and charge transfer in G/BP and BN/BP are similar. Interlayer noncovalent bonding is predicted due to the weak coupling between the $p_z$ orbital of BP and the π orbital of graphene and BN. A small amount of electron transfer from graphene and BN, scaling with the vertical strain, renders BP slightly n-doped for both heterostructures. Several attractive characteristics of BP, including direct band gap and linear dichroism, are preserved. However, a large redistribution of electrostatic potential across the interface is observed, which may significantly renormalize the carrier dynamics and affect the excitonic behavior of BP. Our work suggests that graphene and BN can be used not only as an effective capping layer to protect BP from its structural and chemical degradation while still maintain its major electronic characteristics, but also as an active layer to tune the carrier dynamics and optical properties of BP.

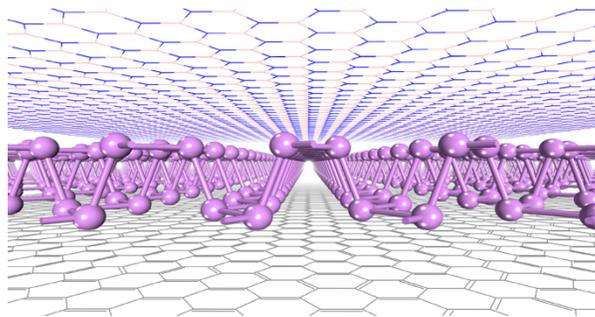

*Supporting Information Placeholder*

## INTRODUCTION

Two-dimensional (2D) atomically thin materials, such as graphene (G), hexagonal boron nitride (BN) and transition metal dichalcogenide (TMD), are of considerable interest owing to their intriguing mechanical, electronic, optical, and electrochemical properties promising for a wide range of applications.[1-4] Recently, the main research focus has shifted from mono-component systems to hybrid ones comprised of at least two types of chemically different 2D materials, such as graphene/BN,[5] graphene/TMD,[6,7] BN/TMD,[8] graphene/silicene,[9,10] and TMD/TMD.[11] These stacked 2D materials form 2D heterostructures, which may possess many intriguing properties, such as novel electronic and optical properties and improved efficiency for carrier injection by modulating the barrier height.[12-16]

More recently, black phosphorene (BP), a new elemental 2D material, has attracted great attention owing to its fascinating properties, such as a finite direct band gap and high mobility (1000 $cm^2V^{-1}s^{-1}$ at room temperature).[17] Importantly, field effect transistors[18-21] and high-frequency nanoelectromechanical resonators[22] based on BP have been demonstrated. Owing to its puckered honeycomb structure, BP shows strongly anisotropic electronic and optical properties,[23-27] which are distinctively different from other 2D materials investigated so far. Monolayer or few-layer BP has been synthesized by mechanical cleavage,[28,29] liquid-phase exfoliation[30] and a transport reaction from red phosphorus.[31] Experimentally, BP/$TiO_2$ hybrid structure[32] for photocatalysis and production of BP quantum dots[33] were reported. Meanwhile, a wealth of theoretical studies were performed to examine its anisotropic electronic structure,[34,35] large excitonic interaction,[36,37] and novel photonic responses.[38,39] First-principles calculations were performed to understand the effects of defects[40,41] and external stimuli, such as strain and electric field,[42-46] on the electronic properties. Interestingly, phosphorene was found to show a giant phononic anisotropy, which is squarely opposite to its electronic counterpart, enabling the asymmetrical transport of electrons/holes and phonons.[47,48]

Despite its many great attractions, a major challenge to use BP for electronic applications is its structural instability at ambient condition.[49,50] Recent works showed that the electronic properties of BP are quite sensitive to external molecules in air[51] and the surface of thin black phosphorus becomes rough after exposure in air within several hours.[52] Nonmagnetic pristine phosphorene becomes magnetic after the adsorption of N, Fe, or Co adatoms, and the magnetic phosphorene with mono-vacancies becomes nonmagnetic after the adsorption of C, N, or Co atoms.[53] Furthermore, spin polarized semiconducting state is realized in phos-phorene by substitutional doping of Ti, Cr, and Mn, while half-metallic state is obtained by V and Fe doping.[54,55] In addition, surface functionalization of phosphorene by different kinds of organic molecules can serve as electron acceptor or electron donor, depending on the relative position of the occupied molecular states to the band edges of phosphorene.[56] Although these interesting studies provide useful insights into controlling the electronic and magnetic properties of phosphorene, the chemical sensitivity of phosphorene to external adsorbates also makes it difficult to achieve robust device performance.

Therefore, enhancing the stability of BP is critically important for any device applications. A recent study showed that the ambient degradation can be suppressed by a proper surface passivation.[57,58] Multilayer BP was predicted to present a better performance than its monolayer counterpart.[34] Besides, few-layer BP flakes passivated by $Al_2O_3$ were found to be stable in ambient air for two months.[59,60] Thus, how to protect phosphorene from structural and chemical degradation is an important research topic. Naturally, chemically stable 2D materials, such as graphene and BN, may be used for protecting fragile, low-chemical stability 2D materials, such as phosphorene and silicene. Indeed, stabilization of silicene by graphene was recently demonstrated,[10] and an improved quality of deposited TMD layers was also found by adopting BN or graphene substrates.[61]

Although it is an attractive idea to use graphene or BN as a capping layer to protect BP, some fundamental issues in the G/BP and BN/BP heterostructures, such as interface bonding, charge transfer, band gap change and band alignment, remain unknown. For device applications based on G/BP and BN/BP, understanding these fundamental issues is highly important and essential. The structural and electronic properties of graphene/phosphorene heterostructure were recently studied.[46] It was found that a perpendicular electric field was able to tune the position of the band structure of phosphorene with respect to that of graphene, which enables a controllable Schottky barrier height in the heterostructure.

In this work, we investigate the electronic structures of two vertically stacked heterostructures, G/BP and BN/BP, by performing first-principles calculations. We find that nonconvalent bonding at the interface and a van der Waals (vdW) gap are formed in both heterostructures. The direct band gap of BP is deserved upon contacting with G or BN owing to the weak coupling of the $p_z$ orbital of BP and the π orbital of graphene and BN. However, a large redistribution of the electrostatic potential across the interface is observed, which may significantly renormalize the carrier dynamics and affect the excitonic behavior of phosphorene. Thus, our work suggests that the graphene and BN can be used not only as a capping layer to protect BP from structural and chemical degradation while still maintain its major electronic characteristics, but also as an active layer to improve the contacting performance of BP with metal electrodes and enable the optimization of the Schottkey barrier by modifying the thickness of graphene or BN layers.

## COMPUTATIONAL METHODS

First-principles calculations are performed by using the plane wave code Vienna *ab initio* simulation package (VASP)[62] within the framework of density functional theory (DFT). Spin-unrestricted generalized-gradient approximation (GGA) with the Perdew-Burke-Ernzerhof functional (PBE) with the projector augmented wave are adopted. vdW-corrected functionals such as Becke88 optimization (optB88),[63] the optimized PBE (optPBE),[64] and the original Dion exchange–correlation functional (vdW-DF)[65] are used to account for the dispersive forces between the atomic sheets. The thickness of the vacuum region is greater than 15 Å to avoid the spurious interaction due to the periodic image. All the structures are optimized until the forces exerted on each atom are less than 0.005 eV/Å. The first Brillouin zone is sampled with a 5×14×1 Monkhorst-Pack grid together with a kinetic energy cutoff of 400 eV. The binding energy ($E_b$) between phosphorene and graphene or BN is calculated as $E_{X/BP}-E_X-E_{BP}$, where $X$ represents graphene (G) or BN, and $E_{X/BP}$, $E_X$ and $E_{BP}$ are the energies of the hybrid bilayer, monolayer graphene or BN, and monolayer phosphorene, respectively.

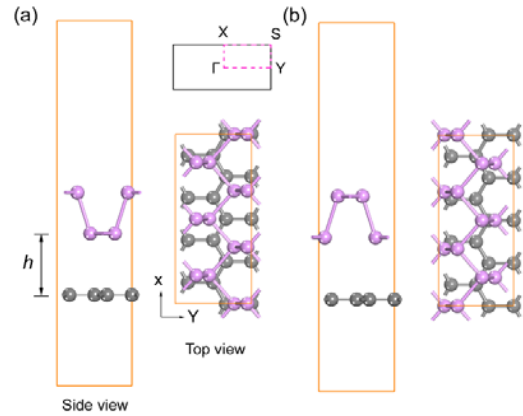

Fig. 1 Lattice structures of the G/BP heterstructure. (a) and (b) show the S1 and S2 structures, respectively. Between them, there is a relative shift of the two monolayers along the lateral direction. The carbon and phosphorus atoms are represented by the grey and violet balls.

## RESULTS

### Structure and energetics of G/BP structure

In order to simulate the electronic properties of the G/BP heterostructure, a bilayer G/BP structure is constructed by combining an orthogonal supercell ($\sqrt{3}$×1) of graphene and a 3×1 supercell of phosphorene. Based on the relaxed structure, the lattice constants of phosphorene are 3.34 and 4.57 Å along the zigzag and armchair direction, respectively. For graphene, the edge length of the basic hexagon is 2.46 Å. Supercells of both layers are created in commensuration with the common lattice, and the lattice constants of the G/BP heterostructure are chosen to be 9.92 and 4.42 Å along the zigzag and armchair direction, respectively, to reduce the strain in both layers. This strategy was also adopted in previous studies for hybrid layers.[9,46] A lattice mismatch of around 3% compressive (tensile) strain occurs along the armchair direction and less than 1% compressive (tensile) strain along the zigzag direction of phosphorene (graphene). Since phosphorene possesses extraordinarily large stretchability along the armchair direction, we therefore also construct G/BP structures by fixing the lattice constant along the armchair direction to that of isolated graphene. The main electronic characteristics include band structure, potential drop and charge transfer between the structures (Note that the charge transfer is calculated by integrating the planar differential charge density as described below). We have adopted these two sets of lattice constants and found that the differences in the bandgap, potential drop and charge transfer are less than 0.1%, suggesting that the differences are negligible.

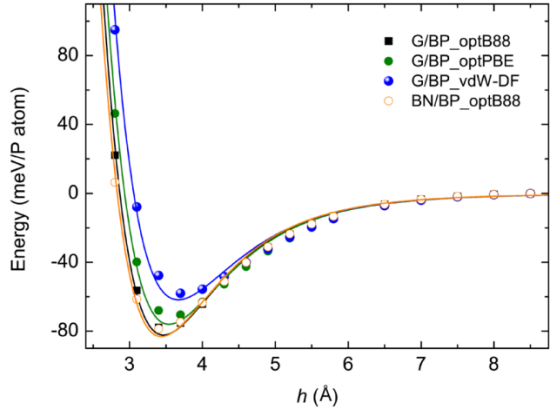

Fig. 2 Interlayer binding energy per phosphorus atom as a function of interlayer distance ($h$) of the G/BP and BN/BP heterostructures. The discrete points represented by the symbols are derived from DFT calculations. The solid lines are the fitting curves based on the Buckingham potential.

To examine the effect of the dispersive forces, we employ three different vdW-corrected exchange-correlation functionals: optB88, optPBE, and vdW-DF. In addition, two different stacking patterns of graphene and phosphorene layers are considered: The bottom ridge of the phosphorene layer directly aligns with the zigzag carbon chain, denoted as S1 structure (Fig.1a); and the bottom ridge of the phosphorene layer aligns with the centers of the hexagons, denoted as S2 structure(Fig.1b).For all the vdW-corrected functionals, the S1 configuration is slightly more stable than the S2 configuration with a slightly larger absolute value of $E_b$ of around 50 meV. Hence, in the following, we mainly consider S1 structure.

Figure 2 shows the $E_b$ as a function of interlayer distance ($h$). The predicted interlayer distance is 3.49, 3.59, and 3.76 Å and the predicted minimum $E_b$ is -79, -71, and -58 meV/P by optB88, optPBE, and vdW-DF functionals, respectively. While all the three functionals are developed based on the non-local vdW functional of Dion et al.,[65] different strategies in describing the exchange and the non-local correlation functional clearly lead to marked differences in quantifying the interlayer interaction. Previous study[66] showed that for graphite, the optPBE give a good prediction for the interlayer binding energy but overestimate the interlayer spacing by ~6%; while the optB88 functional gives a reasonable prediction for both the interlayer distance and binding energy. Therefore, for the G/BP heterostructure, the interlayer spacing is believed to be around 3.49 Å based on optB88 functional and the binding energy $E_b$ is -71 meV/P according to the optPBE functional.

To give an empirical description of the interlayer potentials, the discrete interlayer binding energy per phosphorus atom ($E$) based on optB88 functional calculation as a function of interlayer distance is fitted by the commonly-used Buckingham potential and the Lenard-Jones (LJ) potential:

$$E = ae^{-bh} + c/h^6 \quad (1)$$
$$E = 4\varepsilon[(\sigma/h)^{12} - (\sigma/h)^6] \quad (2)$$

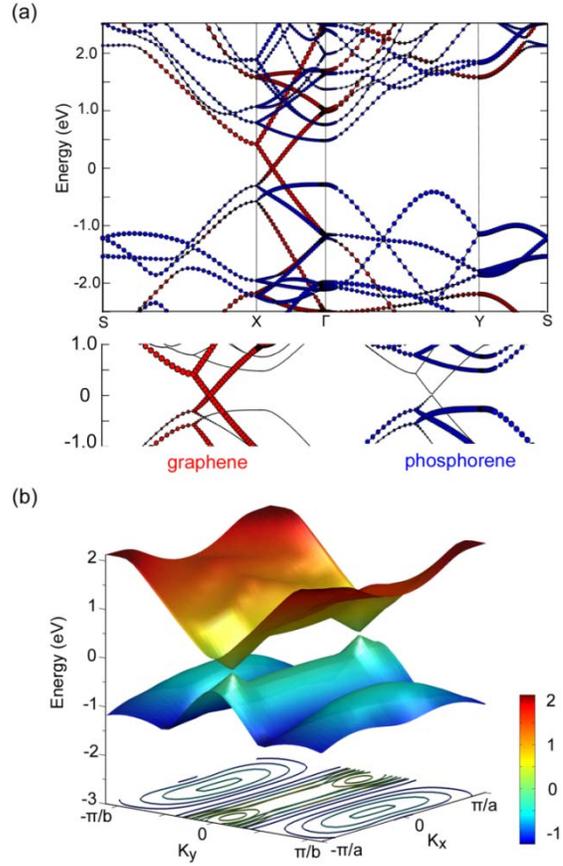

Fig. 3 (a) Electronic band structure of the G/BP heterostructure. The bands projected to graphene and phosphorene orbitals are highlighted by red and blue circles. The circle size reflects the weight of each species in the bands. (b) Isosurface of the valence and conduction bands of the G/BP heterostructure, together with the contour plots of the valence band.

where $a, b, c, \varepsilon, \sigma$ are fitting constants. The first term in the Buckingham potential and LJ potential describes the repulsion and the second one describes the attraction. The fitted values of $a, b, c, \varepsilon,$ and $\sigma$ are $1.184 \times 10^6$ meV, 2.545 Å$^{-1}$, $-4.457 \times 10^5$ meVÅ$^6$, 80 meV, 3.1 Å, respectively.

The interlayer interaction energy, which is obtained by using the vdW-corrected functional, enables the evaluation of the elastic constant in the normal direction ($C_{33}$) of the G/BP heterostructure. Based on the discrete distance-energy points, $C_{33}$ is extracted by calculating the second derivative of the parabolic fitting curve at the energy minimum according to the following formula:

$$C_{33} = \frac{h_0}{S} \frac{\partial^2 E}{\partial h^2} \quad (3)$$

where $h_0$ is the equilibrium interlayer distance at the minimum of the parabola fitting $h$-$E$ curves, $S$ is the area of the supercell, and $E$ is the total energy. The predicted $C_{33}$ of the G/BP heterostructure is 28 GPa from optB88 calculation, which is smaller than that of graphite, 38 GPa.[66,67] The underlying reason of this softening maybe due to the puckered honeycomb structure of phosphorene, which significantly decreases the rigidity, causing a large flexibility along both the in-plane direction and out-of-plane direction.

## Electronic structure of the G/BP heterostructure

The electronic structure of the G/BP heterostructure is shown in Fig. 3. Orbital-decomposed band structure (Fig. 3a) shows that both the Dirac-cone of graphene and the direct band gap of phosphorene are preserved. Due to the breaking of the structural symmetry and the asymmetric potential normal to the layers, there exists a small band gap opening (around 1.3 meV) of graphene in the G/BP heterostructure. Unlike the graphene/silicene bilayer,[9] the Fermi level in the G/BP is located exactly inside the small Dirac bandgap, indicating that the charge doping effect is small in graphene. The bandgap (~0.9 eV, GGA result) of BP layer is nearly unchanged upon contacting with graphene. Note that this band gap value is grossly underestimated due to the well-known deficiency of GGA functional. These results imply that graphene can be a promising candidate either as a capping or supporting layer for encapsulating phosphorene layer or as a sandwiched layer between phosphorene and metal electrode to enhance the contact performance while still maintain the electronic properties of phosphorene.

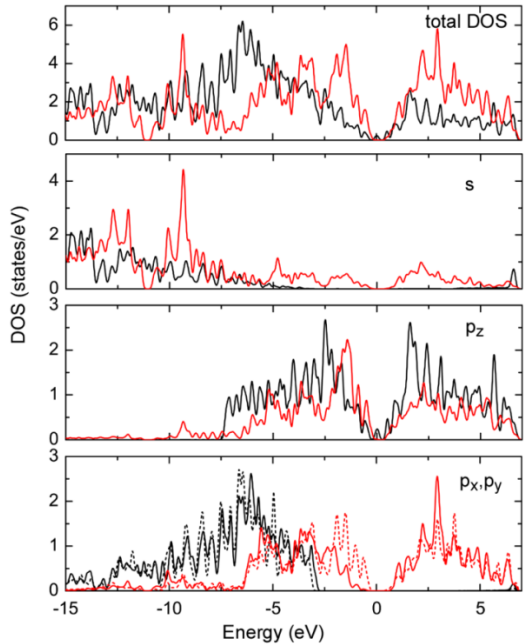

Fig. 4 LDOS of the G/BP heterostructure. The electronic states from phosphorene and graphene are plotted in red and black, respectively. In the bottom panel, the $p_x$ and $p_y$ states are represented by the solid and dashed lines, respectively.

According to Fig. 3a, the valence and conduction bands along the S-X and X-Γ directions show clear orbital hybridization between graphene and BP, while such hybridization is less significant for carriers with momentum along the Y direction (Y-S and Γ-Y paths), which reflects the orientation-dependent interlayer interaction with a stronger orbital coupling along the X direction (the ridge direction of BP sheet) than along the Y direction (the armchair direction of BP sheet). The close-up views of the momentum-energy curves near the Dirac cone are shown in the insets of Fig.3a. Figure 3b shows the surface plots of the valence and conduction bands in the whole 2D Brillouin zone of the G/BP bilayer. Together with the above-analyzed orbital population of the bands as shown in Fig. 3a, it is seen that the long ridge along the Γ-X direction in the surface plot is ascribed to the conduction and valence bands of BP, which are strongly anisotropic, and the Dirac-cone of graphene is directly located above this long ridge (Fig.3b).

The local density of states (LDOS) of the G/BP heterostructure is shown in Fig. 4. For BP, the $s$, $p_x$, $p_y$, $p_z$ states are nearly uniformly distributed across the whole energy spectrum except with in the bandgap, whereas for graphene, the frontier states are only comprised of $p_z$ orbital and there is a gap ranging from -2.5 eV to 5.5 eV for $s$, $p_x$, and $p_y$ orbitals. The striking feature is that in the vicinity of the valence band top, there is only $p_y$ state but no $p_x$ state in phosphorene. Such an asymmetry of the alignment of $p_x$ and $p_y$ orbitals around the Fermi level is critical for the realization of the linear dichroism in phosphorene: Light polarized along the **y** (armchair) direction being strongly adsorbed while that polarized along the **x** (zigzag) being transparent for an energy window between 1.1 and 2.8 eV.[36] The calculated band alignment between graphene and phosphorene shows that the linear dichroism is likely to be preserved in the G/BP heterostructure owing to the transparency of the graphene layer to the linear x- and y- polarized light. This anisotropic character of optical response may be important for identifying the orientation of the capped phosphorene layer below graphene during applications. From the technical viewpoint, identifying the thickness of the bottom BP layer is not straightforward as in pristine BP samples. As the energy window for observing linear dichroism is scaling with the thickness of the BP layer,[23,36] this allows an effective strategy for determining the number of BP layers below graphene.

## Charge redistribution and transfer

We next analyze the charge transfer between the graphene and phosphorene layers. Figure 5(a) shows the differential charge density (DCD) $\Delta\rho(z)$ along the direction normal to the surface, which is calculated by integrating the in-plane DCD. The amount of transferred electron up to $z$ point is given by $\Delta Q(z) = \int_{-\infty}^{z} \Delta\rho(z') \, dz'$. In Fig. 5a, we plot the position-dependent $\Delta\rho(z)$ and $\Delta Q(z)$. It can be seen that the graphene layer donates electrons to the phosphorene layer, which leads to n-doping in phosphorene and p-doping in graphene. The isosurface of the DCD is plotted in Fig.5b, where the loss and the accumulation of electrons across the interface are vividly depicted.

The total gain of electrons in the 4×3 supercell of phosphorene, which is found to be 0.038 e (on average, around 0.003 e per P atom), is determined by the value of $\Delta Q(z)$ at the G/BP interface (defined as the plane of zero charge variation as shown in the $\Delta\rho(z)$ curve). This small charge transfer and weak binding energy indicate that the interaction between the $p_z$ state of BP and the π state of graphene is not strong, and there should be a very small value of the tunneling integral between the $p_z$ state of BP and the $p_z$ orbitals of graphene in the Hamiltonian. Interestingly, our calculations suggest that the interaction can be enhanced upon applying compressive strain normal to the plane. The inset of Fig. 5a shows that the amount of charge transfer increases linearly with decreasing the interlayer distance $h$, ranging from 0.003 e per phosphorus atom at the equilibrium distance of 3.49 Å to 0.005 e at 3.08 Å.

Figure 5c shows the plane averaged electrostatic potential along the perpendicular direction of the G/BP bilayer. The work function of the G/BP bilayer is 4.62 eV. The potential drop ($\Delta V_{G/BP}$) across the bilayer is found to be 6.42 V. Such a large potential difference implies a strong electrostatic field across the interface, which may significantly influence the carrier dynamics and charge injection if the graphene is served as the electrode. In addition, the excitonic behavior of G/BP can be quite different from that of the isolated phosphorene layer as the gradient of the potential across the interface may facilitate the separation of electrons and holes.

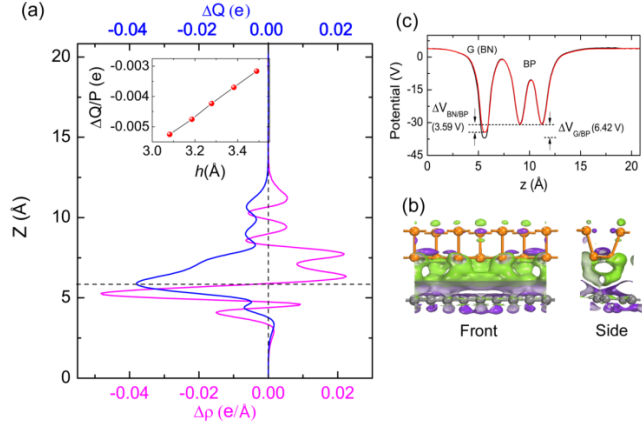

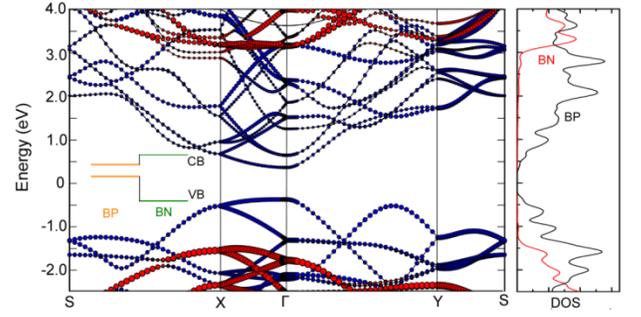

Fig. 5 Charge transfer across the interface of the G/BP heterostructure. (a) The plane-averaged differential charge density $\Delta\rho(z)$ (red) and the amount of transferred charge $\Delta Q(z)$ along the normal direction of the surface (blue). The inset shows the variation of charges per phosphorus atom with the interlayer distance $h$. A negative value signifies a gain of electrons in phosphorus atom. (b) The front and side views of the isosurface of the DCD with the isovalue of 0.005 Å$^{-3}$. The violet (green) color denotes loss (gain) of electrons. (c) Potential profile across the interface for the G/BP (black) and BN/BP(red) heterostructures.

Fig. 6 Band structure (left) and LDOS (right) of the BN/BP heterostructure by HSE06 calculation. The bands projected to BN and phosphorene orbitals are highlighted by red and blue circles. The circle size reflects the weight of each species in the bands. The inset in the left panel shows the type I band alignment of the heterostructure.

### Structural and electronic properties of BN/BP

In the following, we explore the related properties of BN/BP heterostructure. For BN/BP system, the same as G/BP one, the S1 configuration is the most stable structure. Our calculation based on optB88 shows that the equilibrium interlayer distance between BN and BP is 3.46 Å, which is slightly smaller than that of G/BP (3.49 Å). This is similar to the bulk graphite and BN, which are experimentally measured to have essentially the same interlayer distance (~3.33 Å for Bernal stacked graphite and AA' stacked BN) despite their different chemical compositions. As shown in Fig.2, the $E$-$h$ curves of BN/BP are close to those of G/BP and the binding energy $E_b$ is also around -79 meV/P. The interlayer vdW interaction is fitted by using both the Buckingham potential and the LJ potential, and the fitted parameters are shown in Table 1. The similarity in the interlayer binding energy between G/BP and BN/BP is believed to be due to the compensation of electrostatic and dispersive interaction: The more favorable electrostatic interaction in polar BN is balanced by the greater dispersive force in graphene[66] when contacting with phosphorene.

Charge transfer analysis shows that the BN layer acts as a weak donor upon contacting with phosphorene, with each P atom receiving around 0.003 $e$, which is similar to the G/BP case. Figure 6 shows the decomposed band structure of BN/BP. In contrast to G/BP case, the bands associated with phosphorene are closer to the Fermi level. We calculate the work function (6.1 eV) and band gap (5.95 eV) of BN based on HSE06 hybrid functional, together with the previously predicted work function (5.16 eV) and band gap (1.52 eV) of phosphorene,[34] and predict that the band alignment between BN and phosphorene (inset of Fig.6) is a straddling gap (type I). This alignment is also predicted by using PBE calculation, with the values of work function being 5.38 and 4.54 eV, and the values of band gap being 4.90 and 0.35 eV for BN and phosphorene, respectively. The buried electronic states of phosphorene suggest that BN is intrinsically transparent with respect to phosphorene and can be a good protecting layer for phosphorene. However, as shown in Fig. 5c, the potential drop across the BN/BP interface is only 3.59 V, which is much smaller than that of G/BP. Similarly, the electrostatic electric field built at the interface should also exert some effects on the electron-hole recombination in both BN and BP.

## DISCUSSIONS

Phosphorene can be easily doped by external molecules,[51] and thus it is of great interest to explore the effect of intercalated molecules in the G/BP and BN/BP heterostructures. The BN and graphene capping layer should effectively separate BP from those external molecules, and G/BP and BN/BP heterostructures are expected to have a higher chemical stability than the BP monolayer used alone. Moreover, due to the presence of the lone-pair state, monolayer phosphorene can be strongly corrugated when contacting with common metal electrodes, which may degrade its performance and cause additional uncertainties. Owing to the π state and atomically smooth surface, BN and graphene have an excellent structural integrity when contacting with both metal electrodes and phosphorene, thus serving as a perfect sandwiched layer between phosphorene and metal electrode. Moreover, the weak screening behavior in these ultrathin atomic layers suggests that the barrier height can be easily affected by external electric field,[68-70] which opens new possibilities to control the transport of carriers across the interfaces.

In contrast with the widely explored graphene/BN,[5] graphene/TMD,[6,7] BN/TMD,[8] graphene/silicene,[9,10] and TMD/TMD[11]heterostructures, where the two epitaxial layers have the same type of 2D Bravais lattice, the symmetry of the lattice between graphene or BN and BP are different. As a result, the G/BP and BN/BP heterostructures serve as a unique case to explore the effects arising from the asymmetrical potentials with the hexagonal superlattice being superimposed on the orthogonal superlattice. Such integration is expected to induce a novel Moiré pattern and long-range weak periodic potential, thus providing a new approach to tune the direction-dependent transport behavior of electrons and holes in both layers.

Table 1. Best-fit parameters for the interlayer interaction, averaged to each phosphorus atom, of the G/BP and BN/BP heterostructures using the Buckingham and Lenard-Jones (LJ) potentials based on optB88 functional. $C_{33}$ is obtained based on the parabolic curve fitted to the discrete distance-energy points around the equilibrium distance obtained from the optB88 calculations.

|  | $a$ (meV) | $b$ (Å$^{-1}$) | $c$ (meVÅ$^{+6}$) | $\varepsilon$ (meV) | $\sigma$ (Å) | $C_{33}$ (GPa) |
|---|---|---|---|---|---|---|
| G/BP | $1.184\times10^6$ | 2.545 | $-4.457\times10^5$ | 80 | 3.1 | 27.7 |
| BN/BP | $1.137\times10^6$ | 2.544 | $-4.358\times10^5$ | 82 | 3.2 | 23.9 |

## CONCLUSION

We investigate the electronic properties of G/BP and BN/BP heterostructures by using first-principles calculations. Both heterostructures are found to be stabilized by weak vdW interactions and possess a small charge transfer across the interface. The intriguing electronic properties of BP, such as the direct band gap and linear dichroism are maintained upon contacting with graphene or BN, which is an advantage for adopting graphene or BN to cover and protect the chemical- and environment-sensitive phosphorene layer and to enhance its chemical stability and structural integrity in ambient condition. In addition, a large redistribution of electrostatic potential across the interface is observed, which may significantly renormalize the carrier dynamics and affect the excitonic behavior of phosphorene. The potential redistribution across the interface suggests that graphene or BN can serve as an active layer to tune the carrier dynamics of phosphorene. Hence when graphene or BN is inserted in between a metal electrode and phosphorene, such heterostructure can prevent the degradation of phosphorene by avoiding the direct contact of phosphorene with the metal electrode.

## AUTHOR INFORMATION

**Corresponding Author**

zhangg@ihpc.a-star.edu.sg;zhangyw@ihpc.a-star.edu.sg

**Notes**
The authors declare no competing financial interests.

## ACKNOWLEDGMENT

The authors gratefully acknowledge the financial support from the Agency for Science, Technology and Research (A*STAR), Singapore and the use of computing resources at the A*STAR Computational Resource Centre, Singapore.

## REFERENCES

[1] Georgiou,T.; Jalil,R.; Belle,B. D.; Britnell,L.; Gorbachev,R. V.; Morozov, S. V.; Kim,Y.-J.; Gholinia,A.; Haigh,S. J.; Makarovsky,O.; Eaves,L.; Ponomarenko, L. A.; Geim,A. K.; Novoselov,K. S.; Mishchenko, A. Vertical Field-Effect Transistor Based on Graphene-WS$_2$ Heterostructures for Flexible and Transparent Electronics. *Nature Nanotech.* **2013**, *8*, 100-103.

[2] Cai, Y.; Zhang, G.; Zhang,Y.-W. Polarity-Reversed Robust Carrier Mobility in Monolayer MoS$_2$ Nanoribbons. *J. Am. Chem. Soc.* **2014**, *136*, 6269-6275.

[3] Ke,Q.; Tang,C.; Liu, Y.; Liu,H.; Wang,J. Intercalating Graphene with Clusters of Fe$_3$O$_4$ Nanocrystals for Electrochemical Supercapacitors. *Mater. Res. Express* **2014**, *1*, 025015.

[4] Ke, Q.; Liu,Y.; Liu, H.; Zhang,Y.; Hu, Y.; Wang, J. Surfactant-Modified Chemically Reduced Graphene Oxide for Electrochemical Supercapacitors. *RSC Advance* **2014**, *4*, 26398-26406.

[5] Park,S.; Park,C.; Kim,G. Interlayer Coupling Enhancement in Graphene/hexagonal Boron Nitride Heterostructures by Intercalated Defects or Vacancies. *J. Chem. Phys.* **2014**, *140*, 134706.

[6] Yu, W. J.; Liu, Y.; Zhou, H.; Yin, A.; Li, Z.; Huang, Y.; Duan, X. Highly Efficient Gate-Tunable Photocurrent Generation in Vertical Heterostructures of Layered Materials. *Nature Nano.* **2013**, *8*, 952-958.

[7] Wang, Z.; Chen, Q.; Wang, J. Electronic Structure of Twisted Bilayers of Graphene/MoS$_2$ and MoS$_2$/MoS$_2$. *J. Phys. Chem. C* **2015**, DOI: 10.1021/jp507751p.

[8] Huang,Z.; He,C.; Qi, X.; Yang,H.; Liu,W.; Wei,X.; Peng,X.; Zhong,J. Band Structure Engineering of Monolayer MoS$_2$ on h-BN: First-Principles Calculations. *J. Phys. D: Appl. Phys.* **2014**, *47*, 075301.

[9] Hu,W.; Li,Z.; Yang,J. Structural, Electronic, and Optical Properties of Hybrid Silicene and Graphene Nanocomposite. *J. Chem. Phys.* **2013**, *139*, 154704.

[10] Berdiyorov,G. R.;Neek-Amal,M.;Peeters,F. M.; van Duin,Adri C. T. Stabilized Silicene within Bilayer Graphene: A Proposal Based on Molecular Dynamics and Density-Functional Tight-Binding Calculations. *Phys. Rev. B* **2014**, *89*, 024107.

[11] Lu,N.; Guo,H.; Li,L.; Dai,J.; Wang, L.; Mei,W.-N.; Wu,X.;Zeng,X. C. MoS$_2$/MX$_2$Heterobilayers: Bandgap Engineering via Tensile Strain or External Electrical Field. *Nanoscale* **2014**, *6*, 2879.

[12] Geim, A. K.; Grigorieva, I. V. Van der Waals Heterostructures. *Nature* **2013**, *499*, 419-425.

[13] Yin,Z.; Zhang,X.; Cai,Y.; Chen,J.; Wong,J. I.;Tay,Y.-Y.; Chai,J.; Wu,J.;Zeng,Z.;Zheng,B.; Yang,H. Y.; Zhang,H. Preparation of MoS$_2$–MoO$_3$ Hybrid Nanomaterials for Light-Emitting Diodes. *Angew. Chem.* **2014**, *126*, 1-7.

[14] Wang,H.; Liu,F.; Fu,W.; Fang,Z.; Zhou,W.; Liu,Z. Two-Dimensional Heterostructures: Fabrication, Characterization, and Application. *Nanoscale* **2014**, *6*, 12250-12272.

[15] Deng,Y.; Luo,Z.; Conrad,N. J.; Liu,H.; Gong,Y.;Najmaei,S.;Ajayan,P. M.; Lou,J.; Xu,X.; Ye,P. D. Black Phosphorus-Monolayer MoS$_2$ van der Waals Heterojunction p-n Diode. *ACS Nano* **2014**, *8*, 8292-8299.

[16] Yuan,J.;Najmaei,S.; Zhang,Z.; Zhang,J.; Lei,S.;Ajayan,P. M.; Yakobson,B. I.; Lou,J. Photoluminescence Quenching and Charge Transfer in Artificial Heterostacks of Monolayer Transition Metal Dichalcogenides and Few-Layer Black Phosphorus. *ACS Nano* **2015**, *9*, 555-563.

[17] Li,L.; Yu,Y.; Ye, G. J.; Ge, Q.; Ou,X.; Wu,H.; Feng,D.; Chen,X. H.; Zhang,Y. Black Phosphorus Field-Effect Transistors.*Nature Nanotech.* **2014**, *9*, 372-377.

[18] Liu, H.; Neal, A. T.; Zhu, Z.; Luo, Z.; Xu, X.;Tománek, D.; Ye, P. D. Phosphorene: An Unexplored 2D Semiconductor with a High Hole Mobility. *ACS Nano* **2014**, *8*, 4033–4041.

[19] Das,S.;Demarteau,M.;Roelofs,A.Ambipolar Phosphorene Field Effect Transistor. *ACS Nano* **2014**, *8*, 11730–11738.

[20] Liu,H.; Du,Y.; Deng,Y.; Ye,P. D. Semiconducting Black Phosphorus: Synthesis, Transport Properties and Electronic Applications.*Chem. Soc. Rev.* **2015**, DOI: 10.1039/C4CS00257A .

[21] Kamalakar,M. V.;Madhushankar,B. N.;Dankert,A.; Dash,S. P. Low Schottky Barrier Black Phosphorus Field-Effect Devices with Ferromagnetic Tunnel Contacts. *Small* **2015**, DOI: 10.1002/smll.201402900.

[22] Wang,Z.;Jia, H.; Zheng, X.; Yang,R.; Wang,Z.; Ye,G. J.; Chen,X. H.; Shan,J.;Feng,P. X.-L. Black Phosphorus NanoelectromechanicalResonators Vibrating at Very High Frequencies.*Nanoscale* **2015**, *7*, 877-884.

[23] Qiao,J.; Kong,X.; Hu,Z. -X.; Yang,F.; Ji, W. High-Mobility Transport Anisotropy and Linear Dichroism in Few-Layer Black Phosphorus.*Nat. Comm.* **2014**,*5*, 4475.

[24] Xia,F.; Wang,H.;Jia,Y. Rediscovering Black Phosphorus as an Anisotropic Layered Material for Optoelectronics and Electronics.*Nat. Comm.* **2014**, *5*, 4458.

[25] Hong,T.;Chamlagain, B.; Lin,W.; Chuang,H.-J.; Pan, M.; Zhou,Z.; Xu,Y.-Q. Polarized Photocurrent Response in Black Phosphorus Field-Effect Transistors.*Nanoscale* **2014**, *6*, 8978-8983.

[26] Buscema,M.;Groenendijk,D. J.; Steele,G. A.; van der Zant, H. S. J.; Castellanos-Gomez,A. Photovoltaic Effect in Few-Layer Black Phosphorus PN Junctions Defined by Local Electrostatic Gating.*Nat. Comm.* **2014**, *5*, 4651.

[27] Wu,J.; Mao,N.;Xie,L.; Xu,H.; Zhang,J. Identifying the Crystalline Orientation of Black Phosphorus Using Angle-Resolved Polarized Raman Spectroscopy.*Angew. Chem.* **2015**, *127*, 2396-2399.

[28] Lu, W.; Nan, H.; Hong, J.; Chen, Y.; Zhu, C.; Liang, Z.; Ma, X.;Ni, Z.; Jin, C.; Zhang, Z. Plasma-Assisted Fabrication of Monolayer Phosphorene and Its Raman Characterization. *Nano Res.* **2014**, *7*, 853-859.

[29] Castellanos-Gomez,A.;Vicarelli,L.; Prada,E.; Island,J. O.;Narasimha-Acharya,K. L.;Blanter,S. I.;Groenendijk,D. J.;Buscema,M.; Steele,G. A.; Alvarez,J. V.;Zandbergen,H. W.; Palacios,J. J.; van der Zant,H. S. J. Isolation and Characterization of Few-layer Black Phosphorus.*2D Materials* **2014**,*1*,025001.

[30] Yasaei,P.; Kumar,B.;Foroozan,T.; Wang,C.;Asadi,M.;Tuschel,D.;Indacochea, J. E.; Klie,R. F.;Khojin,A. S. High-Quality Black Phosphorus Atomic Layers by Liquid-Phase Exfoliation.*Adv. Mater.* **2015**, DOI:10.1002/adma.201405150.

[31] Köpf,M.; Eckstein, N.; Pfister,D.;Grotz,C.;Krüger,I.;Greiwe,M.; Hansen,T.;Kohlmann,H.;Nilges,T. Access and in Situ Growth of Phosphorene-Precursor Black Phosphorus.*J. Crys. Growth* **2014**, *405*, 6–10.

[32] Lee,H. U.; Lee,S. C.; Won,J.; Son,B.-C.; Choi,S.; Kim,Y.; Park,S. Y.; Kim,H.-S.; Lee,Y.-C.; Lee,J. Stable Semiconductor Black Phosphorus (BP)@Titanium Dioxide (TiO2) Hybrid Photocatalysts. *Sci. Rep.* **2015**, *5*, 8691.

[33] Zhang,X.;Xie,H.; Liu,Z.; Tan,C.; Luo,Z.; Li,H.; Lin,J.; Sun,L.; Chen,W.; Xu,Z.;Xie,L.; Huang,W.; Zhang, H. Black Phosphorus Quantum Dots.*Angew. Chem. Int. Ed.* **2015**, *54*, 1-6.

[34] Cai, Y.; Zhang, G.; Zhang,Y.-W. Layer-Dependent Band Alignment and Work Function of Few-Layer Phosphorene. *Sci. Rep.* **2014**, *4*, 6677.

[35] Li,P.;Appelbaum,I. Electrons and Holes in Phosphorene.*Phys. Rev. B* **2014**, *90*, 115439.

[36] Tran,V.;Soklaski,R.; Liang,Y.; Yang,L. Layer-Controlled Band Gap and Anisotropic Excitons in Few-Layer Black Phosphorus.*Phys. Rev. B* **2014**, *89*, 235319.

[37] Rodin,A. S.;Carvalho,A.; Castro Neto,A. H.Excitons in Anisotropic Two-Dimensional Semiconducting Crystals.*Phys. Rev. B* **2014**, *90*, 075429.

[38] Low,T.; Engel,M.; Steiner,M.;Avouris,P. Origin of Photoresponse in Black Phosphorus Phototransistors.*Phys. Rev. B* **2014**, *90*, 081408(R).

[39] Zhang,S.; Yang, J.; Xu,R.; Wang,F.; Li,W.; Ghufran, M.; Zhang,Y.-W.; Yu, Z.; Zhang,G.; Qin,Q.;Lu,Y. Extraordinary Photoluminescence and Strong Temperature/Angle-Dependent Raman Responses in Few-Layer Phosphorene. *ACS Nano* **2014**, *8*, 9590–9596.

[40] Liu,Y.; Xu,F.; Zhang,Z.;Penev,E. S.; Yakobson,B. I. Two-Dimensional Mono-Elemental Semiconductor with Electronically Inactive Defects: The Case of Phosphorus.*Nano Lett.* **2014**, *14*, 6782-6786.

[41] Dai,J.;Zeng,X. C. Bilayer Phosphorene: Effect of Stacking Order on Bandgap and Its Potential Applications in Thin-Film Solar Cells. *J. Phys. Chem. Lett.* **2014**, *5*, 1289-1293.

[42] Fei,R.;Yang,L. Strain-Engineering the Anisotropic Electrical Conductance of Few-Layer Black Phosphorus. *Nano Lett.* **2014**, *14*, 2884-2889.

[43] Rodin,A. S.;Carvalho,A.;Castro Neto,A. H. Strain-Induced Gap Modification in Black Phosphorus. *Phys. Rev. Lett.* **2015**, *112*, 176801.

[44] Çakır,D.;Sahin,H.;Peeters, F. M. Tuning of the Electronic and Optical Properties of Single-Layer Black Phosphorus by Strain.*Phys. Rev. B* **2014**, *90*, 205421.

[45] Peng,X.; Wei,Q.;Copple,A. Strain-Engineered Direct-Indirect Band Gap Transition and Its Mechanism in Two-Dimensional Phosphorene.*Phys. Rev. B* **2014**, *90*, 085402.

[46] Padilha, J. E.; Fazzio,A.; da Silva, A. J. R. van der Waals Heterostructure of Phosphorene and Graphene: Tuning the Schottky Barrier and Doping by Electrostatic Gating.*Phys. Rev. Lett.* **2015**, *114*, 066803.

[47] Cai,Y.;Ke,Q.;Zhang,G.;Feng,Y. P.;Shenoy,V. B.;Zhang,Y.-W. Giant Phononic Anisotropy and Unusual Anharmonicity

of Phosphorene: Interlayer Coupling and Strain Engineering. *Adv. Funct. Mater.* **2015**, *25*, 2230-2236.

[48] Ong, Z.-Y.; Cai, Y.; Zhang, G.; Zhang, Y.-W. Strong Thermal Transport Anisotropy and Strain Modulation in Single-Layer Phosphorene. *J. Phys. Chem. C* **2014**, *118*, 25272−25277.

[49] Liu, X.; Wood, J. D.; Chen, K.-S.; Cho, E.; Hersam, M. C. In Situ Thermal Decomposition of Exfoliated Two-Dimensional Black Phosphorus. *J. Phys. Chem. Lett.* **2015**, *6*, 773-778.

[50] Island, J. O.; Steele, G. A.; van der Zant, H. S. J.; Castellanos-Gomez, A. Environmental Instability of Few-Layer Black Phosphorus. *2D Mater.* **2015**, *2*, 011002.

[51] Cai, Y.; Ke, Q.; Zhang, G.; Zhang, Y.-W. Energetics, Charge Transfer, and Magnetism of Small Molecules Physisorbed on Phosphorene. *J. Phys. Chem. C* **2015**, *119*, 3102-3110.

[52] Koenig, S. P.; Doganov, R. A.; Schmidt, H.; Castro Neto, A. H.; Özyilmaz, B. Electric Field Effect in Ultrathin Black Phosphorus. *Appl. Phys. Lett.* **2014**, *104*, 103106.

[53] Srivastava, P.; Hembram, K. P. S. S.; Mizuseki, H.; Lee, K.-R.; Han, S. S.; Kim, S. Tuning the Electronic and Magnetic Properties of Phosphorene by Vacancies and Adatoms. *J. Phys. Chem. C* **2015**, *119*, 6530−6538.

[54] Hu, T.; Hong, J. First-Principles Study of Metal Adatom Adsorption on Black Phosphorene. *J. Phys. Chem. C* **2015**, *119*, 8199–8207.

[55] Hashmi, A.; Hong, J. Transition Metal Doped Phosphorene: A First Principles Study. *J. Phys. Chem. C* **2015**, *119*, 9198–9204.

[56] Zhang, R.; Li, B.; Yang, J. A First-principles Study on Electron Donor and Acceptor Molecules Adsorbed on Phosphorene. *J. Phys. Chem. C* **2015**, *119*, 2871–2878.

[57] Wood, J. D.; Wells, S. A.; Jariwala, D.; Chen, K.-S.; Cho, E.; Sangwan, V. K.; Liu, X.; Lauhon, L. J.; Marks, T. J.; Hersam, M. C. Effective Passivation of Exfoliated Black Phosphorus Transistors against Ambient Degradation. *Nano Lett.* **2014**, *14*, 6964-6970.

[58] Gillgren, N.; Wickramaratne, D.; Shi, Y.; Espiritu, T.; Yang, J.; Hu, J.; Wei, J.; Liu, X.; Mao, Z.; Watanabe, K.; Taniguchi, T.; Bockrath, M.; Barlas, Y.; Lake, R. K.; Lau, C. N. Gate Tunable Quantum Oscillations in Air-Stable and High Mobility Few-Layer Phosphorene Heterostructures. *2D Mater.* **2015**, *2*, 011001.

[59] Na, J.; Lee, Y. T.; Lim, J. A.; Hwang, D. K.; Kim, G.-T.; Choi, W. K.; Song, Y.-W. Few-Layer Black Phosphorus Field-Effect Transistors with Reduced Current Fluctuation. *ACS Nano* **2014**, *8*, 11753–11762.

[60] Luo, X.; Hwang, J. C. M.; Liu, H.; Du, Y.; Ye, P. D. Temporal and Thermal Stability of $Al_2O_3$-Passivated Phosphorene MOSFETs. *IEEE Electron Device Lett.* **2014**, *35*, 1314-1316.

[61] Shi, Y.; Zhou, W.; Lu, A.-Y.; Fang, W.; Lee, Y.-H.; Hsu, A. L.; Kim, S. M.; Kim, K. K.; Yang, H. Y.; Li, L.-J.; Idrobo, J.-C.; Kong, J. van der Waals Epitaxy of $MoS_2$ Layers Using Graphene As Growth Templates. *Nano Lett.* **2012**, *12*, 2784−2791.

[62] Kresse, G.; Furthmüller, J. Efficient Iterative Schemes for Ab Initio Total-Energy Calculations Using a Plane-Wave Basis Set. *Phys. Rev. B* **1996**, *54*, 11169.

[63] Becke, A. D. Density-Functional Exchange-Energy Approximation with Correct Asymptotic Behavior. *Phys. Rev. A* **1988**, *38*, 3098.

[64] Klimeš, J.; Bowler, D. R.; Michaelides, A. Chemical Accuracy for the van der Waals Density Functional. *J. Phys. Condens. Matter* **2010**, *22*, 022201.

[65] Dion, M.; Rydberg, H.; Schröder, E.; Langreth, D. C.; Lundqvist, B. I. Van der Waals Density Functional for General Geometries. *Phys. Rev. Lett.* **2004**, *92*, 246401.

[66] Graziano, G.; Klimeš, J.; Fernandez-Alonso, F.; Michaelides, A. Improved Description of Soft Layered Materials with van der Waals Density Functional Theory. *J. Phys. Condens. Matter* **2012**, *24*, 424216.

[67] Chen, X.; Tian, F.; Persson, C.; Duan, W.; Chen, N.-X. Interlayer Interactions in Graphites. *Sci. Rep.* **2013**, *3*, 3046.

[68] Guan, J.; Zhu, Z.; Tománek, D. Tiling Phosphorene. *ACS Nano* **2014**, *8*, 12763-12768.

[69] Zhu, Z.; Tománek, D. Semiconducting Layered Blue Phosphorus: A Computational Study. *Phys. Rev. Lett.* **2014**, *112*, 176802.

[70] Wan, R.; Cao, X.; Guo, J. Simulation of Phosphorene Schottky-Barrier Transistors. *Appl. Phys. Lett.* **2014**, *105*, 163511.